# FLYSIG: Dataflow Oriented Delay-Insensitive Processor for Rapid Prototyping of Signal Processing[1]

**Wolfram Hardt, Bernd Kleinjohann**

e-mail: {hardt,bernd}@c-lab.de

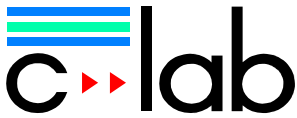

Cooperative Computing & Communication Laboratory
Siemens Nixdorf Informationssysteme AG & Universität-GH Paderborn
Fürstenallee 11, D-33 102 Paderborn, Germany

**Abstract:** *As the one-chip integration of HW-modules designed by different companies becomes more and more popular reliability of a HW-design and evaluation of the timing behavior during the prototype stage are absolutely necessary. One way to guarantee reliability is the use of robust design styles, e.g., delay-insensitivity. For early timing evaluation two aspects must be considered: a) The timing needs to be proportional to technology variations and b) the implemented architecture should be identical for prototype and target. The first can be met also by delay-insensitive implementation. The latter one is the key point. A unified architecture is needed for prototyping **as well as** implementation.*

*Our new approach to rapid prototyping of signal processing tasks is based on a configurable, delay-insensitive implemented processor called FLYSIG[2]. In essence, the FLYSIG processor can be understood as a complex FPGA where the CLBs are substituted by bit-serial operators. In this paper the general concept is detailed and first experimental results are given for demonstration of the main advantages: delay-insensitive design style, direct correspondence between prototyping and target architecture, high performance and reasonable shortening of the design cycle.*

## 1 Introduction

Rapid prototyping for automatically generated designs as well as for manually developed designs has found a lot of interest during the last years [21]. Most approaches map the system's gate-level netlist onto field-programmable gate arrays (FPGAs) mainly due to the reprogramability of the hardware function, that is functionality is easy to change. But in many cases a single FPGA's capacity is not sufficient to cover the complete synthesized netlist and only by additional netlist partitioning an implementation becomes possible [28]. Beside the computation overhead for this partitioning I/O-restrictions must be met [13, 30]. Partitioning and I/O-routing are both highly dependent on the FPGA type, the FPGA interconnections, and the communication protocols. Some providers of multiple FPGA boards offer software for netlist partitioning and generation of communication structures [30, 13]. But these algorithms do not start from an abstract gate-level netlist. The netlist must be mapped onto a concrete gate-library known to the provider, e.g. the LSI10K library [26] and is than automatically re-mapped onto the multiple FPGA board (figure 1).

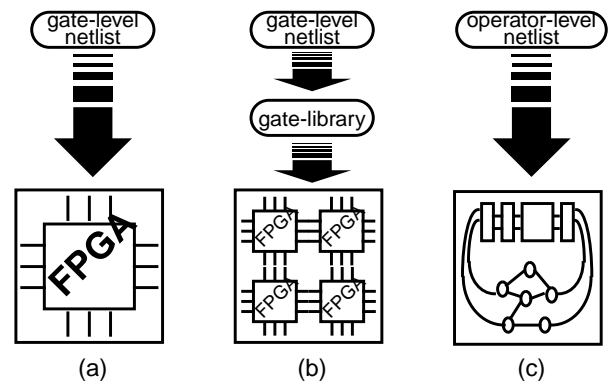

**Figure 1:** Design steps from gate-level netlist to (a) single FPGA, (b) multiple FPGA board and (c) FLYSIG processor based implementations.

In other words, for rapid prototyping the gate-level netlist is mapped to a dedicated FPGA architecture. Thus elements of the netlist are directly decomposed by elements of the FPGA architecture (figure 1 (a)) or by elements of a standard gate library and these elements are decomposed by elements of the FPGA architecture (figure 1 (b)). This double decomposition is the reason for additional costs (number of gate cells, interconnection). It points out that the advantages of FPGA technology are paid by additional design tasks and difficult to meet design restrictions.

We consider an entirely new approach to rapid prototyping solving the mentioned above trials. The main idea is to derive a prototyping architecture from a domain specific optimized target architecture. This architecture is implemented as configurable processor named FLYSIG-prototype processor. The FLYSIG must be once provided as chip, i.e. a new prototyping chip. figure 2 illustrates the comparison of our approach with the described standard approaches.

---

1. The authors would like to acknowledge the support provided by Deutsche Forschungsgemeinschaft DFG, project SPP RP.

2. data**FL**ow oriented dela**Y**-insensitive **SIG**nal processing

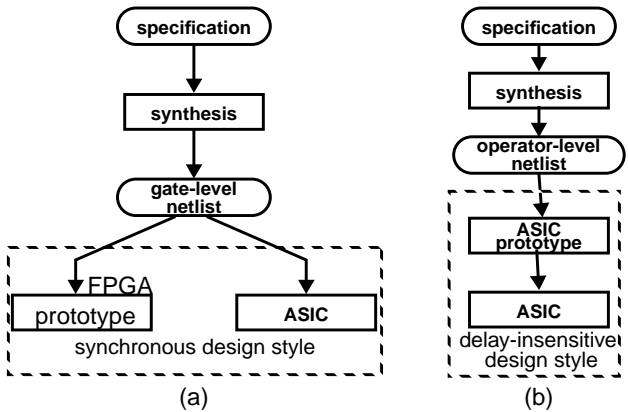

**Figure 2:** Rapid prototyping approaches: (a) synchronous, FPGA based and (b) delay-insensitive FLYSIG-prototype processor based.

The target-architecture itself is specialized to the application domain of fixed digital signal processing algorithms. It is a well known strategy to adapted the design methods to a specific application domain. Different approaches for partitioning and synthesis as well as for target architectures have been proposed, e.g., for control oriented designs [4, 1], data flow oriented designs [14, 3], and real time constrained designs [20, 27]. We applied this principle of specialization to prototyping, i.e., the prototyping-architecture is specialized in respect of the target-architecture. This idea brought us to the FLYSIG-approach. The main advantages are:

- the elimination of all design tasks related to FPGA-prototyping from the design flow. This shortens the design cycle drastically. The additionally introduced design step which derives the FLYSIG-target form the FLYSIG-prototype is an easy to automate task of much lower complexity.

- the delay-insensitive design style used for the FLYSIG-processor. The well known gains of delay-insensitive designs are the elimination of the clock signal, power savings, and a very robust modularization. Delay-insensitivity is of major importance for rapid prototyping because timing analysis on the prototype basis within a complex environment is essential for reliable system validation and short time to market periods. We have examined the synthesis of delay-insensitive modules [16] and found that the timing behavior of such modules can be analyzed in an early design stage, that is the technology impact can be approximated quite well.

- the high performance achieved by the FLYSIG-processor, i.e., sampling rates of 50 MHz and more.

This paper is structured as followed. We present the FLYSIG-processor concept in chapter 3 and illustrate the benefits by the fifths order elliptic filter example in chapter 4. Final conclusions are given in chapter 5. First of all some background information is provided (chapter 2).

## 2 Related work

Research in synchronous design methods has taken place for several decades. A good summary is given in [17]. Basic concepts of asynchronous circuit design are presented e.g. in [15]. A lot of effort has been invested in data protocols and data encoding.

In [12] a true-single-phase protocol has been presented. The two-phase protocol is used by [7] and also the design of the asynchronous version of the ARM processor called AMULET1 [29] is based on the two-phase protocol. In a later version, the AMULET3 processor, the four phase protocol has been used [8] because conversion from two-phase protocol to four-phase protocol is rather costly.

Several data encoding styles are known. Dual-rail encoding provides two single data lines, one for the logic true value and one for the logic false value of a one bit data item. This encoding is rather complex but there are no problems because of hazards [7]. Recently, a combination of single-rail and dual-rail data encoding has been suggested [18]. One approach to reduce the number of data lines necessary for dual-rail encoding is bundled data encoding [22]. To a set of data bits, called bundle, a pair of acknowledge/request bits is added for indication of valid data. Thus the overhead for the bundle is eliminated but the delay of the control lines must be adapted to the delay of the data lines [23]. This limited selection of references shows that a variety of encoding styles and communication protocols have been developed and are used for circuits of reasonable complexity.

Beside data encoding and communication protocols design methodologies have gained a lot of interest. An overview is given in [11]. In 1989 Sutherland presented the concept of micropipelines [25] which has found a lot of interest worldwide. Many investigations have been based on this concept [19, 9, 5, 6, 2]. The concept of multi-ring structures introduced by Staunstrup [24] uses no delay elements but suffer by the complexity of the generated circuits. The performance of multi-ring structures is highly influenced by the availability of data items and free places ready to hand data items on. Free places are commonly called bubbles [10].

In the presented approach, we use the dual-rail data encoding style and the four phase data protocol. We adapted the concept of multi-ring structures and solved the circuit complexity problem by our own ef-ficiently implemented operator library based on the technology described in [16]. Bubbles are integrated in a fixed manner into the operators. Additional bubbles are inserted in between the operators during the design process. The operators are the essential part of the FLYSIG-processor architecture, which is described in the next chapter in some detail.

## 3 The FLYSIG-processor architecture

The FLYSIG-processor architecture allows the efficient implementation of periodic, a priori fixed algorithms. Such algorithms are common practice in digital signal processing, and in real-time controller components, e.g., for reactive robotic systems. All algorithms are also constrained by high sampling rates which are determined by rather complex environments. In this section the architecture itself and the adaptation to prototyping are presented.

### 3.1 Overview

Figure 3 illustrates the dataflow within the FLYSIG-processor which is build out of the three depicted components. In addition interconnection to the environment is provided.

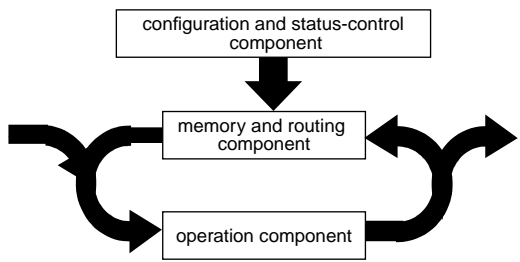

**Figure 3: Dataflow within the FLYSIG-processor.**

The FLYSIG-processor is initialized by the configuration and status-control component. The memory and routing component is interconnected with the operation component to a cyclic structure. Within this structure the multi-ring concept is embedded. Furthermore, this ring-structure is open, that is, additional FLYSIG-processors can be adapted. Thus processor networks can be build easily.

### 3.2 Processor concept

The application which is to be implemented by the FLYSIG-processor is specified by its control/data flow graph. Each operation is represented by a node in the data flow graph. By operation scheduling each dataflow node is assigned to an operator of the operation component of the FLYSIG-processor. Then the interconnection task, known from synchronous design must be performed. For the FLYSIG-prototype version this means to initialize the routing component. For the target-version the chosen routing configuration is hardwired. In figure 4 some details are shown:

(a) The **configuration and status-control component** allows the comfortable configuration of the FLYSIG-processor. The scheduling information is fed into the local memory (registers) and into the routing component for initial operand and result forwarding. In addition the initial operands are stored into the memory. All information can be provided by a configuration host before execution.

(b) The **memory and routing component** handles the operands and the computation results. A data item in the memory is referred to as token, thus it is not distinguished if it is an operand or a result. The tokens flow from the memory into the operation component via the token evaluation. This block determines if a memory cell contains a valid token.

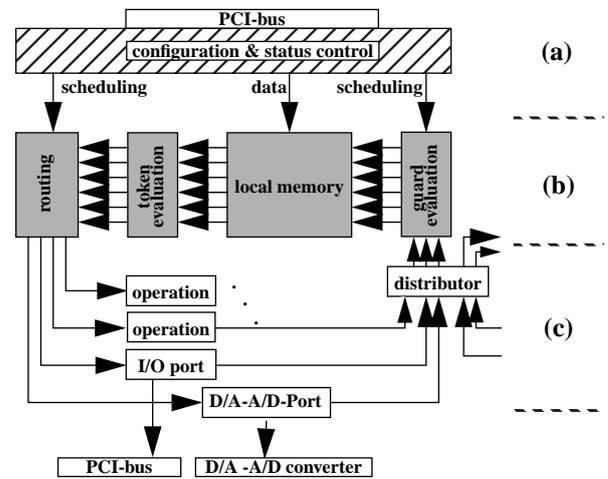

**Figure 4: Concept of the FLYSIG-processor with (a) the configuration and status component, (b) the memory and routing component and (c) the operation component.**

In this case the routing block directs the token to the corresponding operator. Each token consists of the operation id (identifying the operation), a valid-flag segment (indication the availability of operands) and a guard-flag segment (determining where the result values are needed).

(c) The **operation component** implements the operators for all possible computations. In the prototype-version a large set of operators is provided in order to support as many different algorithms as good as possible. Operators read the tokens from memory or from previous involved operators. Thus, operator pipelining is possible. Because of the bit-serial implementation style this leads to deep pipelines with very few hazards and thus to high throughput rates. The computation results are written into the local registers or directly to the registers of a further FLYSIG-processor. This is an important feature which allows to distribute the implementation of a single algorithm over several FLYSIG-processors. Also prototype-versions may be connected with target-versions which are already available. This is of high practical importance because a stepwise migration from the prototype environment to the target implementation becomes feasible.

### 3.3 Prototyping

The prototype version of the FLYSIG-processor differs from the target version only by the implementation of the routing component and the complexity of the operation component. Because the concept of the FLYSIG-prototype version has been derived form the target version the operator concept and the dataflow remains unchanged. Just the hardwired scheduling implementation is exchanged by a configurable one. This allows the mapping of several different algorithms onto the same FLYSIG-prototype processor. In this context a configurable scheduling can be implemented by simple associatively controlled crossbar switches. Furthermore, in the FLYSIG-prototype processor a set of operators is provided with most common operators. This operator set is only restricted by the design size. Once an algorithm has been mapped onto the prototype version and hardware-in-the-loop simulation has been successful the FLYSIG-target processor can be derived from the prototype version easily by eliminating all unused operators and replacing the configurable scheduling by a hardwired implementation. This eliminations and replacements can be performed automatically.

### 3.4 Operators

The FLYSIG's operation component provides operators with control, storage elements and arithmetic functionality.

#### 3.4.1 Control operators

For control of delay-insensitive multi-ring based architectures several operators have been presented by Staunstrup [24] including asymmetric switches, join, and fork operators. We extended this set of control operators by select operators which allow the communication between different rings. The block symbol and the register-transfer level netlist of the read-select operator are presented in figure 5.

The RSELECT operator is controlled by the diamond input which determines from which input the data is read. The opposite behavior is realized by the WSELET operator reading form the only input and writing to the indicated output. Both operators are essential to implement control flow.

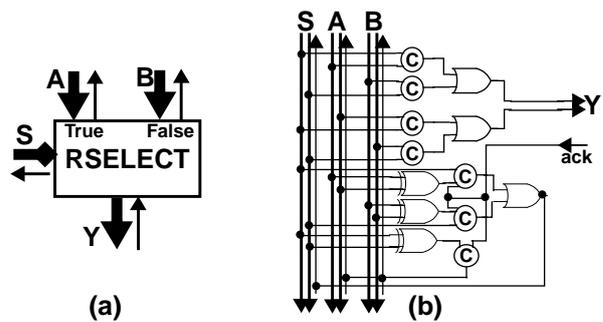

**Figure 5: a) Block symbol and (b) RT-netlist of read-select operator.**

#### 3.4.2 Storage elements

Three basic register types are needed. All are derived from an uninitialized minimal register. In addition a 0-initialized register and a 1-initialized register is needed. These basic register elements can be queued to shift registers. Is is important that for each data bit within the shift register an extra empty basic register element should be provided thus optimal throughput can be reached.

#### 3.4.3 Fixpoint arithmetic operators

Operators for fixpoint arithmetic can be constructed out of c-gates and synchronous *OR* gates. Such circuits can be generated by standard two level synthesis technics whereby the *AND*-plane is substituted by a c-gate plane [24]. This design style is called DIMS[1] and leads to a mixture of synchronous and asynchronous gate-level components and employs large numbers of c-gates.

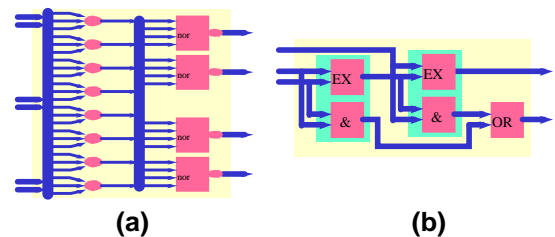

**Figure 6: Add-operator: (a) DIMS and (b) complete dual-rail implementation.**

We build operators based on dual-rail compatible implementation of logic gates. This eliminates the large number of c-gates and ensures a completely time invariant design. In figure 6 both implementations are compared. The complexity of a c-gate (circle) and a dual-rail *AND*-gate is comparable.

By this bit-serial add-operator and several basic register elements a complete bit-serial full-adder can be constructed. The RT-level netlist is given in figure 7.

---

1. **D**elay/**I**nsensitive **M**in-term **S**ynthesis

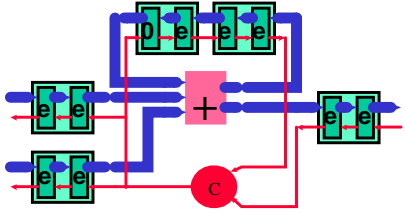

**Figure 7: Complete bit-serial full-adder netlist.**

The modularity is quite obvious and because of the dual-rail data encoding delay-insensitivity is maintained on each hierarchy level. By this operator implementation style further operators are implemented and simulated on RT-level. Simulation is based on a VHDL behavioral description of each basic entity. Detailed timing data from transistor-netlist simulation is used within the RT-level VHDL descriptions which allows very fast realistic evaluation of timing.

### 3.4.4 Optimization

The implementation style for FLYSIG-operators allows high optimizations for the implementation of operation queues. For illustration, we consider the computation for the term $x' = a + x + x + x$. A straight forward implementation is shown in figure 8 (a). Three full-adders and three basic registers are allocated.

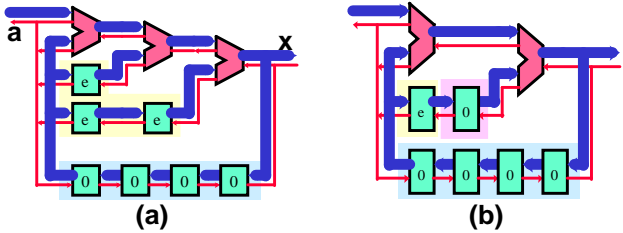

**Figure 8: (a) Straight forward and (b) optimized implementation of the term x'=a+x+x+x**

A much cheaper solution with exactly the same functionality is given in figure 8 (b). Only two full-adders and two basic register elements are needed, whereby one register element has been initialized. This inserted data item implements a shift operator with very low costs.

## 4  Example

For demonstration of the FLYSIG-processor's concept and benefits we present an example. It is taken from the well known high level synthesis benchmark suit. The fifths order elliptic filter requires reasonable computation performance and is simple enough for demonstration. From this filter smaller subcomponents have been derived for detailed case studies.

| example | # operations | # registers | #feedback cycles |
|---|---|---|---|
| elliptic_filter | 26 | 8 | 8 |
| filter_ab 1 | 3 | 3 | 3 |
| filter_ab 2 | 3 | 2 | 3 |
| filter_ab 3 | 3 | 1 | 3 |
| filter_abc_1 | 6 | 3 | 3 |
| filter_abc_2 | 6 | 2 | 3 |
| filter_abc_3 | 6 | 1 | 3 |
| filter_abcd_1 | 9 | 3 | 3 |
| filter_abcd_2 | 9 | 2 | 3 |
| filter_abcd_3 | 9 | 1 | 3 |

**Table 1. Characteristics of filter benchmarks**

All filter benchmarks have been mapped onto the FLYSIG-prototype processor. Based on this prototype timing evaluation has been performed. As a first step we have implemented a VHDL environment for simulation of the FLYSIG-prototype processor. This includes VHDL descriptions of all gate-level cells, operational units and complete operators. The timing characteristics of a the used gate level cells have been obtained from analog simulation of the transistor netlists and where imported in the VHDL implementations. Based on this two level simulation realistic timing and functional evaluation can be performed very quickly. All simulations up to several thousand ns execution time of the FLYSIG-processor could be performed within some cpu seconds which is negligible few compared to other approaches e.g., based on petri-net simulation or single transition graph simulation.

Considering the timing characteristics, the circuit's latency and throughput are important. We have both examined for the FLYSIG-implementations of the filter applications of table 1. Latency is determined by the longest operational path within the circuit. In addition the number of registers is important because registers are used to implement bubbles. Thus higher latency values are found for the same filter-functionality implemented with fewer registers. Of course, this is a performance/size trade-off. The determined latency values are depicted in figure 9. But throughput is of much higher importance because latency can be regarded as system setup time. In figure 10 the best throughputs for all examined filter applications are shown which are reached by an optimal number of bubble registers.

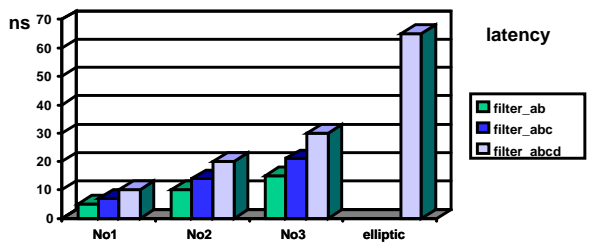

**Figure 9: Latency of filter applications implemented by FLYSIG-operators.**

All filters show the same throughput rate although the number of operations differ. This is due to the deeply pipelined operators and the delay-insensitive design style. The throughput is only restricted by the operator's throughput which is rather high because of the efficient implementation of the dual-rail gates.

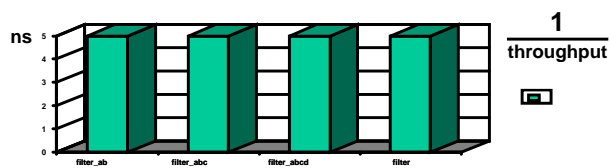

**Figure 10: Throughput of filter applications implemented by FLYSIG-operators.**

## 5 Conclusion

In this paper we presented a new methodology for rapid prototyping of cyclic signal processing applications. The FLYSIG processor was developed for prototyping. From the FLYSIG-prototype implementation the Flysig-target can be derived easily. It has been shown on simulation base that this prototyping methodology provides very fast prototype environments wherein hardware-in-the-loop simulation is possible.

Further investigations will include the extension of the operator library by floating-point operators as well as by trigonomic operators. The automation of all design tasks specialized to the FLYSIG- processor is also under development.

### References

[1] R.A. Bergamaschi, D. Lobo, and A. Kuehlmann. Control optimization in High-Level Synthesis using Behavioral Don't Cares. In *Proc. of the 29th DAC*, pages 657–661. ACM/IEEE, 1992.

[2] Kees van Berkel and Arjan Bink. Single-track handshaking signaling with application to micropipelines and handshake circuits. In *Proc. International Symposium on Advanced Research in Asynchronous Circuits and Systems*, pages 122–133. IEEE Computer Society Press, March 1996.

[3] J.C. Bier. DSP Processors and Cores: The Optios Multiply. In *Integrated System Design*, pages 56 – 67, June 1995.

[4] K. Buchenrieder and C. Veith. A Prototyping Environment for Control-Oriented HW/SW Systems useing State-Charts, Activity-Charts, and FPGAs. In *Proc. of the EDAC*. IEEE, 1994.

[5] Chih-Ming Chang and Shih-Lien Lu. Performance issues on micropipelines. *IEEE Technical Committee on Computer Architecture Newsletter*, October 1995.

[6] Paul Day and J. Viv Woods. Investigation into micropipeline latch design styles. *IEEE Transactions on VLSI Systems*, 3(2):264–272, June 1995.

[7] Mark Dean, Ted Williams, and David Dill. Efficient self-timing with level-encoded 2-phase dual-rail (LEDR). In Carlo H. S'equin, editor, *Advanced Research in VLSI: Proceedings of the 1991 UC Santa Cruz Conference*, pages 55–70. MIT Press, 1991.

[8] S. B. Furber, P. Day, J. D. Garside, N. C. Paver, and S. Temple. AMULET2e. In C. Muller-Schloer, F. Geerinckx, B. Stanford-Smith, and R. van Riet, editors, *Embedded Microprocessor Systems*, September 1996. Proceedings of EMSYS'96 - OMI Sixth Annual Conference.

[9] S. B. Furber and O. A. Petlin. Scan testing of micropipelines. In *Proc. IEEE VLSI Test Symposium*, pages 296–301, May 1995.

[10] Mark R. Greenstreet and Kenneth Steiglitz. Bubbles can make self-timed pipelines fast. *Journal of VLSI Signal Processing*, 2(3):139–148, November 1990.

[11] Scott Hauck. Asynchronous design methodologies: An overview. Technical Report TR 93-05-07, Department of Computer Science and Engineering, University of Washington, Seattle, 1993.

[12] Hong-Yi Huang, Kuo-Hsing Cheng, et. al. Low-voltage low-power CMOS true-single-phase clocking scheme with locally asynchronous logic circuits. In *Proc. International Symposium on Circuits and Systems*, pages 1572–1575, 1995.

[13] Quickturn Systems Inc. *Emulation System User's Guide*, release 4.4 edition, 1993.

[14] Aswaree Kalavade and Edward A. Lee. Hardware-Software Codesign Methodology for DSP Applications. *IEEE Design & Test of Computers*, pages 16–28, September 1993.

[15] M. Kishinevsky, Lavagno L., and Vanbekbergen P. Tutorial: The Systematic Design of Asynchronous Circuits. Technical report, Proc. of the ICCAD, 1995.

[16] B. Kleinjohann. *Synthese von zeitinvarianten Hardware Modulen*. PhD thesis, University of Paderborn, 1994. Dissertation.

[17] Luciano Lavagno and Alberto Sangiovanni-Vincentelli. *Algorithms for Synthesis and Testing of Asynchronous Circuits*. Kluwer Academic Publishers, 1993.

[18] Gensoh Matsubara and Nobuhiro Ide. A low power zero-overhead self-timed division and square root unit combining a single-rail static circuit with a dual-rail dynamic circuit. In *Proc. International Symposium on Advanced Research in Asynchronous Circuits and Systems*. IEEE Computer Society Press, April 1997.

[19] R. Mehra and J. D. Garside. A cache line fill circuit for a micropipelined, asynchronous microprocessor. *IEEE Technical Committee on Computer Architecture Newsletter*, October 1995.

[20] A. Moitra and M. Joseph. Implementing real-time systems by transformation. In H. Zedan, editor, *In Real-time Systems: Theory and Applications*. North Holland, 1990.

[21] S. Note, P. van Lierop, and van Ginderdeuren. Rapid Prototyping of DSP systems: requirements and solutions. In *Sixth IEEE International Workshop on Rapid System Prototyping*, pages 88–96, Chapel Hill, North Carolina, USA, June 1995.

[22] Ad Peeters and Kees van Berkel. Single-rail handshake circuits. In *Asynchronous Design Methodologies*, pages 53–62. IEEE Computer Society Press, May 1995.

[23] Per Torstein Røine. Building fast bundled data circuits with a specialized standard cell library. In *Proc. International Symposium on Advanced Research in Asynchronous Circuits and Systems*, pages 134–143, November 1994.

[24] Jens Sparsø, Jørgen Staunstrup, and Michael Dantzer-Sørensen. Design of delay insensitive circuits using multi-ring structures. In *Proc. EURO-DAC*, pages 15–20, Hamburg, Germany, 1992.

[25] Ivan E. Sutherland. MICROPIPELINES. *Communications of the ACM*, 32:720–738, June 1989.

[26] Synopsys, Inc., Mountain View, CA. *VHDL Design Compiler (tm) Manual*, 3.3a edition, 1995.

[27] J. Vanhoof, K.V. Rompaey, I. Bolsens, G. Goosens, and H. De Man. *High-Level Synthesis for Real-Time Digital Signal Processing*. Kluwer Academic Publishers, Boston/Dordrecht/London, 1993.

[28] M. Wendling and W. Rosenstiel. A Hardware Environment for Prototyping and Partitioning Based on Multiple FPGAs. In *Proc. of the EDAC*, pages 77–82, Grenoble, France, 1994. IEEE.

[29] J. V. Woods, P. Day, S. B. Furber, J. D. Garside, N. C. Paver, and S. Temple. AMULET1: An asynchronous ARM processor. *IEEE Transactions on Computers*, 46(4):385–398, April 1997.

[30] Zycad Corporation, Inc., USA. *Concept Silicon Software (tm) Manual*, 6.0 edition, 1994.